\documentclass[12pt,preprint]{aastex}
\def\lea{\mathrel{<\kern-1.0em\lower0.9ex\hbox{$\sim$}}}
\def\gea{\mathrel{>\kern-1.0em\lower0.9ex\hbox{$\sim$}}}

\slugcomment{Submitted to The Astronomical Journal}

\shorttitle{UVOT Stars: M~67}
\shortauthors{Siegel et al.}

\begin{document}

\title{The {\it Swift\/} UVOT Stars Survey. I. Methods and Test Clusters}

\author{Michael H. Siegel\altaffilmark{1},
Blair L. Porterfield\altaffilmark{1},
Jacquelyn S. Linevsky\altaffilmark{1,2},
Howard E. Bond\altaffilmark{1,3,7},
Stephen T. Holland\altaffilmark{3},
Erik A. Hoversten\altaffilmark{1,4}, 
Joshua L. Berrier\altaffilmark{1},
Alice A. Breeveld\altaffilmark{5},
Peter J. Brown\altaffilmark{6},
Caryl A. Gronwall\altaffilmark{1}
}

\altaffiltext{1}{The Pennsylvania State University, Department of Astronomy and
Astrophysics, 525 Davey Laboratory, University Park, PA, 16802
(siegel@astro.psu.edu, blp14@psu.edu, heb11@psu.edu, caryl@astro.psu.edu)}

\altaffiltext{2}{Cypress Bay High School, 18600 Vista Park Blvd  Weston, FL
33332}

\altaffiltext{3}{Space Telescope Science Institute, 3700 San Martin Drive,
Baltimore, MD, 21218 (sholland@stsci.edu)}

\altaffiltext{4}{Current Address: Department of Physics and Astronomy,
University of North Carolina at Chapel Hill, Phillips Hall, CB 3255, 120 E.
Cameron Ave., Chapel Hill, NC 27599-3255 (ehoverst@live.unc.edu)}

\altaffiltext{5}{Mullard Space Science Laboratory, University College London,
Holmbury St. Mary, Dorking, Surrey RH5 6NT (aab@mssl.ucl.ac.uk)}

\altaffiltext{6}{George P. and Cynthia Woods Mitchell Institute for Fundamental
Physics \& Astronomy,  Texas A. \& M. University, Department of Physics and
Astronomy, 4242 TAMU, College Station, TX 77843, USA
(grbpeter@yahoo.com)}       

\altaffiltext{7}{Visiting Astronomer, Kitt Peak National Observatory, National
Optical Astronomy Observatory, which is operated by the Association of
Universities for Research in Astronomy, Inc., under cooperative agreement with
the National Science Foundation.}

\begin{abstract}

We describe the motivations and background of a large survey of nearby stellar
populations using the Ultraviolet Optical Telescope (UVOT) aboard the {\it Swift
Gamma-Ray Burst Mission}.  UVOT, with its wide field, NUV sensitivity, and
2\farcs3 spatial resolution, is uniquely suited to studying nearby stellar
populations and providing insight into the NUV properties of hot stars and the
contribution of those stars to the integrated light of more distant stellar
populations.  We review the state of UV stellar photometry, outline the survey,
and address problems specific to wide- and crowded-field UVOT photometry. We
present color-magnitude diagrams of the nearby open clusters M~67, NGC~188, and
NGC~2539, and the globular cluster M~79.  We demonstrate that UVOT can easily
discern the young- and intermediate-age main sequences, blue stragglers, and hot
white dwarfs, producing results consistent with previous studies.  We also find
that it characterizes the blue horizontal branch of M~79 and easily identifies a
known post-asymptotic giant branch star.

\end{abstract}


\section{Introduction: Stellar Populations in the UV}
\label{s:intro}

At far-ultraviolet (FUV) and near-ultraviolet (NUV) wavelengths---accessible
only from high altitude or from space---old stellar populations are dominated by
hot stars in short-lived stages of stellar evolution. In optical images of these
old populations, such hot objects are inconspicuous compared to the large
numbers of main-sequence stars and red giants. By contrast, young populations are dominated
by luminous massive stars whose spectral-energy distributions also peak in the
UV\null.  Thus UV imaging of resolved populations provides a unique window into
stellar properties and evolution.

Over the past four decades, UV space missions have imaged globular and open star
clusters and other resolved populations, in order to obtain better insight into
the UV photometric properties of hot stars.  These experiments have included the
{\it Orbiting Astronomical Observatory 2\/} ({\it OAO-2}), {\it Astronomical
Netherlands Satellite\/} ({\it ANS}), {\it Ultraviolet Imaging Telescope\/}
({\it UIT}), {\it Galaxy Evolution Explorer\/} ({\it GALEX}), and {\it Hubble
Space Telescope\/} ({\it HST})\null. The present paper focuses on the most
recently launched spacecraft with a UV imaging camera, the {\it Swift Gamma-Ray
Burst Mission\/} with its Ultraviolet Optical Telescope (UVOT)\null.  The
capabilities of these missions for UV imaging, and a sample of studies that have
used them to focus on stellar populations, are summarized in
Table~\ref{t:uvmissions}. General reviews of UV studies of stellar populations
include Kondo et al.\ (1989), O'Connell et al.\ (1999), Bowyer et al.\ (2000),
Moehler et al.\ (2001), Catelan (2009), and Heber (2009).

\begin{deluxetable}{lcccccl}
\rotate
\tabletypesize{\scriptsize}
\tablewidth{0 pt}
\tablecaption{Space-Based FUV/NUV Imaging of Globular Clusters\label{t:uvmissions}}
\tablehead{
\colhead{Mission} &
\colhead{Mission} &
\colhead{Aperture} &
\colhead{FOV Area} &
\colhead{Resolution} &
\colhead{Wavelength} &
\colhead{Star-Cluster Study Examples}\\
\colhead{} &
\colhead{Duration} &
\colhead{(cm)} &
\colhead{(arcmin$^2$)} &
\colhead{($''$)} &
\colhead{Range (\AA)} &
\colhead{}}
\startdata
{\it OAO-2}     & 1968--1973 &\nodata  & 78.5 & \nodata  & 1000--4250   & Molnar et al.\ 1978\\
                &            &         &      &	       &  	     & de Boer \& Code 1981\\
{\it ANS}       & 1974--1976 & 22   & 6.25     & \nodata  & 1500--3300   & van Albada et al.\ 1979\\
{\it UIT}       & 1990, 1995&30   & 1257     & 2.7      & 1200--3200   & Parise et al.\ 1994; Whitney et al.\ 1994, 1995\\
          & &     &          &	       &  	     & Hill et al.\ 1996; Dorman et al.\ 1997\\
          & &     &          &	       &  	     & O'Connell et al.\ 1997; Parise et al.\ 1998\\
          & &     &          &	       &  	     & Landsman et al.\ 1996, 1998\\
{\it GALEX}     & 2003--2013 & 50   & 4071     & 4.3--5.3  & 1350--2800   & Lanzoni et al.\ 2007; Dalessandro et al.\ 2009, 2012\\
          & &     &          &	       &             & Schiavon et al.\ 2012\\
{\it HST}/FOC   & 1990--1999 & 240  & .01--.05  & 0.01-0.04      & 1100--6500   & Ferraro \& Paresce 1993; Burgarella et al.\ 1994\\
{\it HST}/WFPC2 & 1993--2009 & 240  & 2.3--5.0  & .04      & 1150--11000  & Ferraro et al.\ 1999, 2001; Castellani et al.\ 2006\\
          & &     &          &          &             & Sandquist et al.\ 2010; Brown et al.\ 2010a\\
          & &     &          &          &             & Dalessandro et al.\ 2011, 2013a\\
{\it HST}/ACS   & 2002-- & 240  & 0.21--.30 & .03      & 1150--11000  & Dieball et al.\ 2007; Haurberg et al.\ 2010\\
{\it HST}/WFC3  & 2009-- & 240  & 7.3      & .04      & 2000--10000  & Bellini et al.\ 2013; Piotto et al.\ 2013\\
{\it Swift}/UVOT & 2004--      & 30   & 289      & 2.3      & 1700--8000   & This study\\
\enddata
\end{deluxetable}

The UV light of stellar populations is contributed by the following stellar
types:

\begin{enumerate}

\item Massive Main-Sequence Stars: Main-sequence stars of spectral type F and
earlier have significant UV emission.  The properties of the most massive O- and
B-type stars and the track of their post-main-sequence evolution are still
poorly understood owing to their short lifetimes, uncertain distance scales, and
the poorly constrained UV extinction law (see reviews in Zinnecker \& Yorke
2007, Portegies Zwart et al.\ 2010, and Langer 2012).  

\item Luminous Blue Variables and Wolf-Rayet Stars: The evolutionary stages
between the O-star main sequence and supernovae are highly extremely luminous
and have peak emission in the UV.  They are also highly variable (Langer 2012).

\item Population I Supergiants: Evolved high-mass stars of spectral types late B
through early G are the optically brightest members of young populations (e.g.,
Humphreys 1983) because of their small bolometric corrections compared to blue
and red supergiants (e.g., Bond 2005).  They are of particular interest in the
UV as they have large Balmer discontinuities in their spectra owing to low
surface gravities, which can make them stand out on optical-UV color-color
diagrams.

\item Blue Stragglers (BSS): Primarily present in older stellar populations,
these are merged binaries that mimic the colors and magnitudes of higher-mass
stars.  They are a particularly important contributor to the UV light of
globular clusters (Laget et al.\ 1992; Mould et al.\ 1996; Ferraro et al.\ 2001,
2003; Dalessandro et al.\ 2013b).

\item Blue Horizontal-Branch (BHB) Stars: BHB stars are He-burning stars lying
blueward of the RR~Lyrae gap; they are found in most old, metal-poor stellar
populations.  They typically have temperatures of 7,000--16,000~K and therefore
have significant UV emission.  UV color-magnitude diagrams (CMDs) have shown
discontinuites in the HB morphology at approximately 11,500 K and 21,000
K\null.  These HB ``jumps" are created, respectively, by radiative levitation of
heavy elements in the atmospheres of hot evolved stars (Grundahl et al.\ 1999)
and early helium flashing (Momany et al.\ 2004).

\item Extreme Horizontal Branch Stars (EHB): EHB stars or hot OB subdwarfs are
helium-burning HB stars having such thin hydrogen envelopes that they cannot
sustain shell burning.  These stars will eventually evolve directly to the white
dwarf (WD) sequence without ascending the asymptotic giant branch (AGB)\null. 
With temperatures greater than 16,000~K, they are extremely luminous in the UV
and likely produce much of the ``UV upturn" seen in old elliptical galaxies
(Brown et al.\ 1997; O'Connell et al.\ 1999), possibly dominating the UV light
of old stellar populations.  This feature has been identified in numerous
Galactic globular clusters (e.g., Rich et al.\ 1997; Sandquist et al.\ 2008; Brown et al.\ 2010a;
Schiavon et al.\ 2012).

\item Blue-Hook Stars (Bhk): Bhk stars are horizontal-branch stars that appear
to be below the He-burning limit and hotter than the theoretical limit of EHB
stars ($\sim$35,000~K).  Only discovered in the most massive globular clusters,
these may be very rare ``late flashers" that ignite helium while on the
WD sequence (D'Cruz et al.\ 2000; Dalessandro et al.\ 2011; Brown et
al.\ 2012) and have high surface helium abundances (Moehler et al.\ 2004a,b,
2007).

\item AGB Manqu\'e Stars (AGB-M): These failed AGB stars are the descendants of
EHB and Bhk stars. They evolve away from the He-burning sequences but have
envelopes too small to reach the AGB (Greggio \& Renzini 1990).

\item Post-Asymptotic Giant Branch Stars (PAGB): These stars are rapidly
evolving from the tip of the AGB to the WD sequence.  They are the most
luminous Population~II stars, and are the brightest stars in globular clusters
in the optical and UV (Strom et al.\ 1970; Zinn et al.\ 1972).  Their rapid
evolution results in extremely short lifetimes ($\sim$25,000 yr).  This makes
them both rare and valuable as they may have a uniform optical luminosity while 
at spectral types A and F, making them potential standard candles (Bond 1997;
Alves et al.\ 2001).

\item Hot White Dwarfs: The degenerate end states of low-mass stars, these start
at temperatures of over 100,000 K and cool adiabatically.  In the early stages,
they are extremely bright in the UV. Our previous studies have explored some of
their photometric properties (Siegel et al.\ 2010, Siegel et al.\ 2012).

\end{enumerate}

All of the stars listed above are detectable in the optical and infrared. 
However, with the exception of BHB, MS and supergiant stars, they are often drowned in the
light of the far more numerous late-type MS and RGB stars.  Although they are
hot, the bolometric corrections make them  optically faint  and
indistinguishable from cooler stars.  And for very hot stars ($T_{\rm
eff}>10,000$~K), the optical passbands are far enough down the Rayleigh-Jeans
tail of the spectral-energy distributions that the stars all have similar
optical colors over a wide range of temperatures. In the UV, by contrast, the
numerous late-type stars emit very little light.  Therefore, the rarer hot
stars---EHB, Bhk, AGB-M, PAGB, etc.---are easily discerned in the NUV and
FUV\null.  Illustrations of this can be seen in Figure 4 of Dieball et al.\
(2007),  Figure 2 of Haurberg et al.\ (2010), and Figure 4 of Brown et al.\
(2010a), which contrast the unusual and rarefied topography of FUV/NUV CMDs
against the more familiar optical.

Star clusters have been the primary resource for identifying and characterizing
hot stars.  Star clusters contain simple (mostly) stellar  populations and are
therefore ideal for constraining the UV properties of stellar populations as a
function of age and metallicity.  They are nearby, meaning they can be studied
on a star-by-star basis.  And most clusters are well-studied in the optical and
infrared, meaning the underlying age, metallicity, distance, and foreground
reddening are well known.  The Galaxy contains more than 800 open clusters,
which are primarily young, and  150 globular clusters, which are primarily old. 
Together, they span a broad range of stellar populations in both age and
chemistry, as well as a variety of environments.

Despite the extensive, excellent and pioneering work cited above, frustrating
gaps remain in our understanding of the UV properties of stellar populations and
hot stars. Examples include the role that helium abundance plays in creating EHB
and Bhk stars, the luminosity of high-mass stars, and the reality of the
``late-flasher" scenario for creating EHB and AGB-M stars.  Hanging over all of
these questions is the uncertainty in the UV extinction law, particularly the
strength of the 2175~\AA\ bump and the shape of the UV extinction curve at
shorter wavelengths (see Hoversten et al., in prep).  Studies of open clusters,
as opposed to globular clusters, are particularly lacking with concomitant lack
of insight into the UV properties of young stellar populations and massive
stars.  And measures of the integrated UV light of stellar populations,
particularly young stellar populations, are needed to provide a critical check
on the population-synthesis models used to measure the stellar content of
distant unresolved stellar populations.  Bruzual (2009), discussing the PEGASE
galaxy evolution code, noted that ``The main limitations of these models come
from incomplete data sets of evolutionary tracks and from either empirical or
theoretical stellar spectral libraries." More effort is needed to produce
complete censuses of hot stars in nearby stellar populations;  connect the
number, type, luminosity and colors of those stars to the properties of the
underlying stellar populations and/or individual stars; understand the shape of
the UV extinction curve; and, perhaps most significantly, produce measures of
integrated light to compare to more distant and unresolved stellar populations. 
This need is particularly critical for young stellar populations.

The core problem is a lack of data.  Hot stars are rare and require extensive
surveys to identify.   Moreover, work beyond the optical window requires space
instrumentation.  However, even the existing space instruments have been unable
to provide a complete census of hot stars.  The {\it HST\/} surveys of simple
stellar populations have produced spectacular color-color and CMDs, but lack the
spatial footprint for a complete census.  {\it GALEX\/} has provided photometry
of many stars but lacks the spatial resolution to provide high-quality
photometry in globular clusters and its brightness constraints limited its
ability to study open clusters.  UIT made significant
contributions but the data are more than two decades old and not deep enough
to characterize the fainter populations.

To address these lacunae, we have initiated a series of programs using the UVOT
onboard the {\it Swift Gamma-Ray Burst Mission\/} (Gehrels et al.\ 2004; Roming
et al.\ 2005).  UVOT occupies a unique niche that makes it ideal for studying
nearby stellar populations in the NUV\null. Its field of view covers 40 times
the area of {\it HST}/WFC3, and it has twice the spatial resolution of {\it
GALEX}\null. Importantly, UVOT has less stringent brightness limits than {\it
GALEX}, allowing it to survey UV-bright areas of the sky including the Galactic
mid-plane, open and globular clusters, and the Magellanic clouds, all of which
are now being surveyed by {\it Swift\/}.

Our first effort to use {\it Swift}/UVOT to constrain the properties of hot
stars created a set of UV standard stars and demonstrated an ability to
constrain the temperatures of WDs to a precision of $\sim$1000~K, based on Sloan
Digital Sky Survey, UVOT, and {\it GALEX\/} data (Siegel et al.\ 2010). 
Subsequently, we demonstrated UVOT's ability to detect and characterize hot
pre-WD stars, clearly identifying a hot compact companion to the red-giant
central star in the planetary nebula WeBo~1 (Siegel et al.\ 2012).  We are now
undertaking the next step in characterizing hot stars in the UVOT photometric
system: constraining the properties of hot main-sequence and evolved stars in
nearby open and globular clusters. This will inform future studies of other
galaxies within the Local Group.

We begin this series of papers here by describing our methods of wide-field
photometry and the connection between the measured photometry and the properties
of the underlying stars.  \S2 details the data, processing and analysis
methods, and the conversion of raw UVOT instrumental magnitudes to calibrated
photometry.  It is likely of interest only to those attempting to perform wide-field photometry
with UVOT. \S3 then compares the properties of three open clusters---M~67,
NGC~188, and NGC~2539---to theoretical isochrones, attempting to map out the
color-magnitude and color-color space occupied by the primary sequences.  It
also examines the UVOT photometry of the old globular cluster M~79.  \S4
summarizes our results.  In an appendix, we address the effect of reddening on
UVOT photometry.

\section{UVOT Observing and Data Reduction}
\label{s:obsred}

\subsection{The Swift Ultraviolet Optical Telescope (UVOT)}
\label{ss:uvotintro}

The {\it Swift\/} spacecraft has a complement of three instruments, used
respectively at gamma-ray, X-ray, and UV\slash optical wavelengths, with a 
primary mission of detecting and studying gamma-ray bursts and their X-ray and
UV\slash optical afterglows.  UVOT is a modified Ritchey-Chr\'etien 30 cm
telescope that has a wide ($17' \times 17'$) field of view and a microchannel
plate intensified CCD detector operating in photon-counting mode (see details in
Roming et al.\ 2000, 2004, 2005).  Incoming photons are amplified by a photomultiplier
stage creating an electron cloud.  This cloud is converted back into photons by a phosphor
screen and the splash of those photons is recorded and centroided
on a fast-read 256$\times$256 CCD to one-eighth of a pixel precision.
The camera is equipped with a 11-slot filter wheel that includes a clear white
filter, $u$, $b$, and $v$ optical filters, $uvw1$, $uvm2$, and $uvw2$ UV
filters, a magnifier, two grisms, and a blocked filter.
The $uvw2$ and $uvw1$ filters have substantial
red leaks, which have been characterized to high precision by Breeveld et al.\ (2010) and are included
in the current UVOT filter curves (this issue is discused in detail by Brown et al\ (2010b) and Siegel et al.\ (2012)).
Table \ref{t:swiftfilt} list the central wavelengths and FWHM of UVOT's imaging filters. It also lists the
strength of the red tail, defined as the ratio of the integrated effective area
curve beyond 3000 \AA\ over the integrated effective area curve from 1600--7000 \AA. The observational
effect of the red tails depends on the color of the source being observed.  Table 12 of Brown et al.\ (2010b) indicates
that the effect of the red leak is small for stars with effective temperatures greater than 10,000 K.  The effect
of the red leak is included in all isochrones used in this study.cd

\begin{deluxetable}{cccc}
\tablewidth{0 pt}
\tablecaption{Swift UVOT Filters\label{t:swiftfilt}}
\tablehead{
\colhead{Filter} &
\colhead{Central Wavelength} &
\colhead{FWHM} &
\colhead{Red Sensitivity}\\
\colhead{} &
\colhead{(\AA)} &
\colhead{(\AA)} &
\colhead{}}
\startdata
\hline
$v$    & 5468 & 769 & \nodata \\
$b$    & 4392 & 974 & \nodata \\
$u$    & 3465 & 785 & \nodata \\
$uvw1$ & 2600 & 693 & 0.11\\
$uvm2$ & 2246 & 498 & 0.0019\\
$uvw2$ & 1928 & 657 & 0.024\\
\hline
\enddata
\end{deluxetable}

UVOT's wide field, 2\farcs3 resolution, broad wavelength range (1700--8000 \AA),
photometric stability, and ability to observe simultaneously with {\it Swift}'s
X-Ray Telescope (XRT; Burrows et al.\ 2005) allow a broad range of scientific
investigations, over and beyond the gamma-ray follow-up. UVOT is especially well
suited, in the context of hot stars, for studying nearby star clusters.  Its
field of view can enclose almost any nearby cluster in a single pointing, and
its spatial resolution allows photometry of stars almost to the center of the
most crowded fields.  In this sense, it is similar to UIT, which had a
wide-field solar-blind channel that enabled some of the early space-based
studies of UV-bright stars and stellar populations.

\subsection{Photometric Reduction of UVOT Images}
\label{ss:uvot}

In this section, we describe how we process UVOT data to produce science-quality
images. A general description of current UVOT data processing is given in the
online UVOT
Digest\footnote{\url{http://heasarc.gsfc.nasa.gov/docs/swift/analysis\slash
uvot\_digest.html/}}. UVOT data are processed with programs contained in the
HEASARC FTOOLS software
package\footnote{\url{http://heasarc.gsfc.nasa.gov/docs/software/lheasoft/}},
which includes standard pipeline processes, custom-built UVOT image tools, and
up-to-date calibration files. Here we emphasize several important steps that are
relevant to our program of wide-field stellar photometry.     Our procedure for
producing images free of instrumental signatures involves five steps: data
acquisition, aspect correction, exposure-map generation, large-scale sensitivity
correction, and image combination.

\begin{enumerate}

\item Data Acquisition: Raw and pipeline-processed UVOT data are downloaded
directly from the HEASARC
archive\footnote{\url{http://heasarc.gsfc.nasa.gov/cgi-bin/W3Browse/swift.pl/}}.
(The same processed UVOT data are also available through the MAST archive at
STScI\footnote{http://archive.stsci.edu}.)  The pipeline will be described in
detail in Marshall et al.\ (in preparation). The standard download includes the
raw frames, processed images that have been geometrically corrected to sky
coordinates, and exposure maps.  We also download auxiliary spacecraft data for
reasons given below.

\item Aspect Correction: The HEASARC automated processing performs a
transformation of the raw CCD images to celestial coordinates, using the
spacecraft attitude files (not the image-header coordinates, which are for the
{\it nominal\/} pointings rather than the {\it actual\/} pointings).  The
reduction pipeline then runs UVOTASPCORR to match stars in the frame to the
USNO-B1 astrometric catalog for fine aspect correction. However, this automated
aspect correction is prone to failure in some fields.  The severe crowding, and
the sometimes dramatic differences in UV and optical fluxes for late-type stars,
can cause this process to fail or (more rarely) produce an inaccurate solution. 
This is most likely to occur in globular clusters and fields in the Magellanic
Clouds.

When the automatic fine aspect correction fails, we manually match stars in each
frame to reference images from the STScI Digitized Sky Survey.  We then use the
FTOOLS program UVOTUNICORR to generate a new aspect solution, which is applied
to the image if the new scatter in offsets is less than 1\farcs25.  We have
found that about ten matched stars are needed to produce a sound
two-dimensional translation.  The FTOOL UVOTEXPMAP is then used to regenerate
the exposure map at the new coordinates.

\item Exposure Map Regeneration: The exposure maps are a critical part of any
analysis of UVOT data.  A typical {\it Swift\/} observing window is 600--1800~s
long.  Observations longer than about 1000~s will likely be  spread over
multiple spacecraft orbits. In our program, the observations have often been
spread over weeks, months, or even years. Thus the observations have been taken
at a range of spacecraft roll angles.  There is also a range in pointings, at a
level of 1--$2'$. These factors cause the total exposure time in a combined
image of a target to vary significantly over the field, as shown in the left
panel of Figure \ref{f:flatfield}.

\begin{figure}
\epsscale{0.75}
\plotone{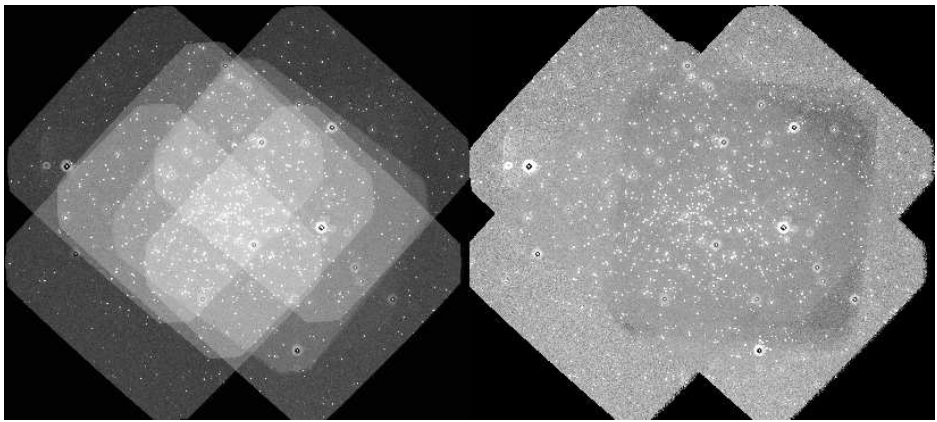}
\caption{{\it Left:} stacked deep image of NGC~188,  created by combining
processed UVOT data. Total exposure varies across the image. {\it Right:}
combined image divided by the UVOT exposure map.  This count-rate image is more
uniform but retains some residual variation due to varying background level from
the Earth limb.\label{f:flatfield}}
\end{figure}

The exposure map images generated by the HEASARC pipeline are, absent aspect
correction problems, perfectly suited to this task.  However, we regenerate them
in all cases so that we can mask out the outer regions of the UVOT images, where
the data quality is lower. The first step in this process is to generate a
bad-pixel map for each image, using the UVOTBADPIX tool and the standard
bad-pixel map in the most recent calibration database (CALDB)
build\footnote{\url{https://heasarc.gsfc.nasa.gov/docs/heasarc/caldb/swift/}}.
We then recreate the exposure map using UVOTEXPMAP with the spacecraft attitude
files and a simple 25-pixel trim to generate masks that remove the outskirts of
each image. This step is automatically taken when we have to revise the
automatic aspect correction for an image. This trim is more aggressive than used
in our standard products or the forthcoming UVOT source catalog (Yershov et al.,
in preparation), but was chosen to avoid false detections in the image corners.

\item Large Scale Sensitivity (LLS): UVOT data are not corrected in the standard
pipeline for flat-field effects.  We make this correction using a sensitivity
map, as described in Breeveld et al.\ (2010). This task grows more complicated
when stacking multiple images at varying pointings and roll angles.  The first
step in correcting for the flat field is to generate individual sensitivity
images for each exposure.  This task is performed using the FTOOLS task
UVOTSKYLSS, the LSS maps available in CALDB, and the spacecraft attitude files.

\item Image Combination: The sensitivity maps, exposure maps, and images are
then combined into single large multi-extension fits files using the FAPPEND
task.  UVOTIMSUM is then used to combine these into a single image, a single
exposure map, and a single LSS map in each filter, adjusting the combination
method as appropriate for each type of image. In each case, we use the mask
files generated during the exposure map reconstruction to produce clean masked
images, removing both bad pixels and chip edges.

\end{enumerate}

\subsection{Wide Field Stellar Photometry with UVOT}
\label{ss:photometry}

UVOT's photon-counting mode and photon-cloud centroiding of counts complicates
the analysis of wide-field photometry in a number of ways. The problems of
coincidence and dead-time losses, in which two or more photons hit a similar
detector location during the same read time, or a photon hits during the
readout, are addressed in Poole et al.\ (2008) and Breeveld et al.\ (2010).
However, there are additional secondary effects that can impact PSF photometry.
The onboard centroiding of the photon ``splashes" causes the coincidence loss at
any pixel to be mixed into adjacent pixels.  This causes the PSF to change shape
at high count rates, becoming narrower and developing a darkened halo of lower
count rates where photons in the outskirts of the PSF are lost (although the
overall counts are conserved, so the Poole et al.\  coincidence loss corrections
will still work for fixed aperture photometry). In addition, the clusters we are
studying are crowded, with multiple overlapping point sources that complicate
both the individual photometric measures and the corrections for coincidence
loss.

UVOT photometry is generally performed with the FTOOL UVOTSOURCE, which provides
single-source aperture photometry, or UVOTDETECT, which incorporates the Source
Extractor program (Bertin \& Arnouts 1996) to provide multi-source photometry.  
While these tools have their advantages, including automatic correction for
coincidence loss, they are not well-suited for images as crowded as in our
cluster program fields. Thus we have adapted the DAOPHOT/ALLSTAR PSF photometry
program (Stetson 1987, 1994) to the analysis of UVOT data. Ideally, software
unique to UVOT would be developed that accounts for the change of PSF shape with
count rate.  However, as shown below, a PSF independent of count rate has proven
capable of producing high-quality photometry, providing one selects the PSF
stars carefully.

We first aperture photometer all stars using a circular aperture of radius
5\farcs0.  We then carefully select PSF stars that are not near image edges,
bright stars, or the ``burned in" areas of the cluster centers. We also have
found that, because UVOT is a non-linear detector, removing the bright stars
from our PSF sample produces significant improvements in the object subtraction
and overall photometry. These stars are most obvious when the coincidence loss
produces visible artifacts---dark halos and boxes around bright stars.  However,
we have found that an aggressive trimming of the bright end of the PSF stars was
necessary to produce high-quality photometry.  We allowed the PSF to have
quadratic variation over the image, although in fact the PSF is reasonably
constant over the face of the detector.  Once this PSF star selection stage is
completed, we follow standard methods of stellar photometry using the ALLSTAR
program.

Translating the instrumental DAOPHOT photometry to calibrated magnitudes,
however, involves a number of intermediate steps.  These steps are included in
the builds of UVOTSOURCE and UVOTDETECT\null.   We list these here as they are
necessary for any photometric method that assumes linear count rates.

\begin{enumerate}

\item We convert the DAOPHOT instrumental magnitudes and sky measures to
raw count rates.

\item The count rates are corrected for exposure time using the summed
exposure map.

\item We aperture-correct the PSF photometry to the standard 5\farcs0 aperture
used for calibration.  It is critical that this step be done before the
coincidence-loss correction or the correction will fail to linearize the
photometry.  We have found that DAOPHOT sometimes produces a small zero-point
offset between the aperture photometry and the PSF photometry.  This is
non-trivial with UVOT since the count rates are not linear.  Correcting the
photometry back to the 5\farcs0 apertures before the coincidence loss correction
removes the non-linearity.

\item The count rates---total count rate in the aperture (star plus sky)
and sky count rate---are corrected for coincidence loss using the formulation
of Poole et al.\ (2008).  We then subtract the corrected sky rate from the
corrected total count rate to produce a coincidence-loss corrected count rate
for each star.

\item The count rates are then corrected for the large-scale sensitivity and the
secular decline in instrument sensitivity, using the large-scale sensitivity map
and formulations from Breeveld et al.\ (2010), respectively.

\item Finally, we convert the count rates back to stellar magnitudes and apply
the zero-point corrections of Breeveld et al.\ (2011). Zero-points are available
for both Vega-based magnitudes and AB~magnitudes.

\end{enumerate}

\begin{figure}[h]
\begin{center}
\includegraphics[scale=0.9]{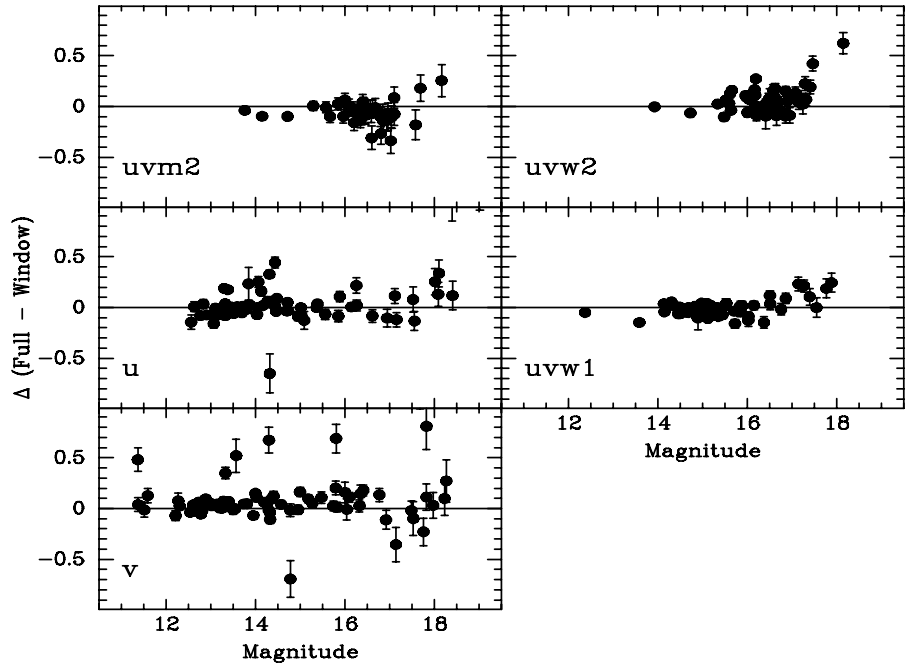}
\end{center}
\caption{A comparison of {\it Swift}/UVOT DAOPHOT photometry from deep
full-frame images to shallow windowed images of the open cluster M~67. Despite
the differences in readout characteristics, the comparison is linear in all five
observed passbands. There are slight variations in zero point.  The comparison
is in Vega magnitudes.\label{f:windowcomp}}
\end{figure}

To test the internal consistency of the pipeline, we compared deep
photometry of the open cluster M~67 (described in \S \ref{s:results}) to photometry
from shallower images taken in
windowed mode.  The windowed mode, which images the inner $5'\times 5'$ region
of the UVOT field, significantly shortens the read and dead times of the array,
reducing the coincidence losses for bright targets.  Figure \ref{f:windowcomp}
shows the comparison for five of the six UVOT filters (no windowed data were
obtained in the $b$ filter).  In all five cases, fits to weighted residuals show
that the comparisons are flat to
within 1--2\% and statistically consistent with a flat trend, confirming the accuracy of the Poole et al.\ (2008)
coincidence-loss corrections (the comparisons appear non-linear at the faint end due to poorly measured stars being
preferentially detected when they scatter to brighter magnitudes in the windowed frame). 
There are slight offsets in the zero points
(ranging from about 0.01 to 0.06 mag, with a mean of 0.008~mag). These are
barely statistically significant, and are likely the result of poor sky
statistics on the short windowed images. There is also some scatter at the faint end ($<$ 17 mag) from marginal detections
when stars in the deep images being scattered to brighter magnitudes in the shallow images.

\clearpage

\subsection{Comparison with Ground-Based Optical Photometry}
\label{ss:groundcomp}

As a test of the external accuracy of our photometry, we compare our UVOT data
for M~67 with calibrated ground-based observations of the same stars. Frames of
M~67 were obtained by H.E.B. with the Kitt Peak National Observatory (KPNO) 0.9
m telescope in 1998 March and 1999 January and March, using the T2KA chip with
exposure times ranging from 5 to 120~s. The KPNO frames were reduced using the
IRAF\footnote{IRAF footnote.} CCDPROC task, and instrumental stellar magnitudes
were measured with DAOPHOT/ALLSTAR and DAOGROW (Stetson 1990). The photometry
was then corrected for atmospheric extinction, and calibrated to the Johnson
{\it BV\/} system using observations on the same nights of standard-star fields
of Landolt (1992) and matrix-inversion methods described in Siegel (2002). We also
compare our photometry to the published $BV$ photometry of Yadav et al.\ (2008), obtained
with the ESO Wide-Field Imager and calibrated to the secondary standards of Stetson (2000, 2005), which is also
tied to the Landolt system.  The
UVOT photometry has been transformed to the Johnson system using the formulations of
Poole et al.\ (2008).

\begin{figure}[h]
\begin{center}
\includegraphics[scale=0.9]{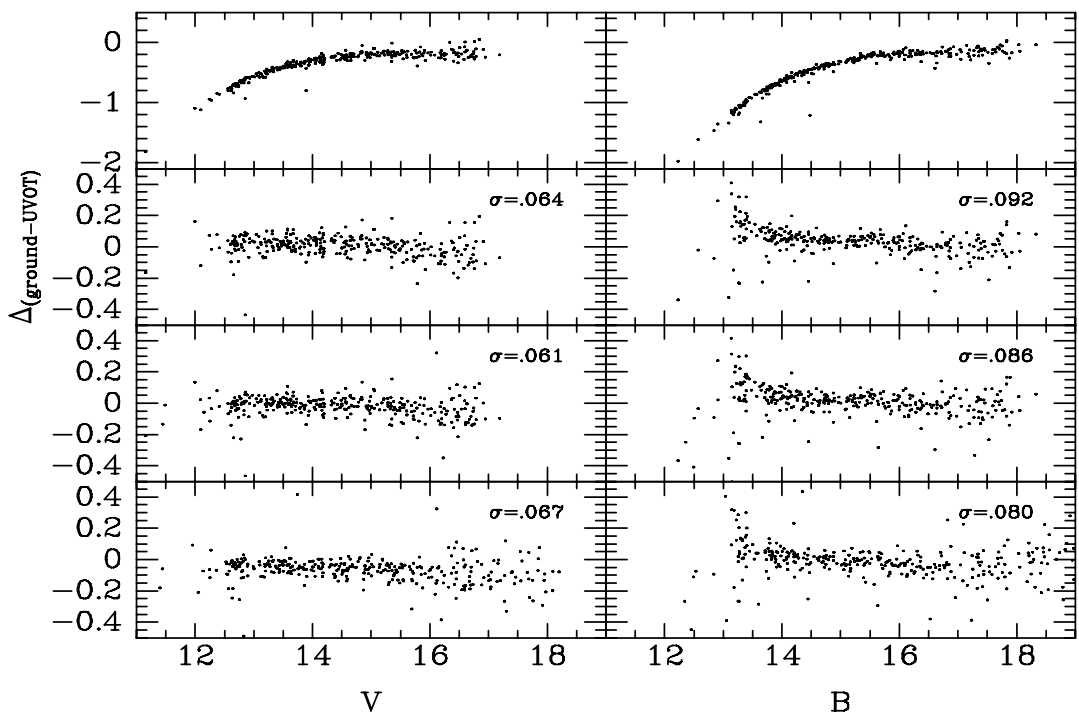}
\end{center}
\caption{A comparison of {\it Swift}/UVOT DAOPHOT photometry to ground-based
photometry for the open cluster M~67.  The top panels show the raw photometry.
The strong non-linearity is the result of UVOT's coincidence loss.  The middle
panels are corrected for coincidence loss but not for the large-scale
sensitivity.  The non-linearity is mostly removed but some scatter remains.
The bottom two panels are corrected for all instrumental effects, with the third
panel comparing to unpublished photometry from KPNO and the bottom comparing
the published ESO photometry of Yadav et al.\ (2008). The comparison is
in Vega magnitudes.\label{f:gbcomp}}
\end{figure}

Figure \ref{f:gbcomp} compares the UVOT and ground-based photometry for both $V$
and $B$ magnitudes.  The top panels show the photometric comparison to the KPNO data with no
correction of the UVOT data for coincidence loss or large-scale sensitivity; 
only zero-point offsets and sensitivity decline were incorporated. As can be
seen, the uncorrected UVOT photometry severely underestimates the brightnesses
of the brightest stars.

In the second two panels the coincidence-loss corrections have been applied. Now
the non-linearity has mostly disappeared.  However, there is still significant scatter, which our analysis shows
is dependent on position within the frame.  These geometric residuals are a bit
haphazard as they follow the complex overlap of three different UVOT
observations.

The bottom four panels show the comparisons with the large-scale sensitivity
corrections also applied, with the third set comparing to the KPNO photometry
and the fourth comparing to ESO photometry.  The geometric and zero point residuals are now
reduced and we find the weighted fit to be linear within 1\%.  Large outliers
are likely extended objects or blends.  We note that while
the scatter in the comparison plots is somewhat large, this represent a worst-case scenario for
UVOT because the exposures times are short ($\sim 160s$).  Our full survey uses far deeper
imaging, usually 1-2 ks.

\clearpage

\section{UVOT UV and Optical Photometry of Four Star Clusters}
\label{s:results}

In this section, we examine the CMDs of the old open clusters M~67 and NGC~188,
the young cluster NGC~2539, and the old globular cluster M~79, based on archival
{\it Swift}/UVOT data.  Details of the observations are listed in Table
\ref{t:swiftphot}, including total exposure time and photometric depth. Note
that, because of the aforementioned variation of exposure time over the face of
the image and the effect of crowding in the cluster cores, the exposure times
and limiting magnitudes represent the {\it maximum\/} $3\sigma$ depth in each field.  The
effective depth is shallower in outlying regions, which are not as well-exposed,
and in the cluster cores, where the background diffuse light of unresolved stars
is bright.

These data were obtained as follows: (1)~M~67 was observed on 2009 March 29 and
2011 January 11 as a calibration target, and again on 2011 May 30 as part of
M.H.S.'s {\it Swift\/} fill-in program.  (2)~NGC~188 was extensively observed in
a broad mosaic from 2007 December 2 to 6 as a calibration target to probe the 
large-scale sensitivity of UVOT\null. Additional data were taken on  2007
October 31  and December 20. (3)~NGC~2539 was observed from 2011 June 18 to 19
as part of M.H.S.'s fill-in program.  (4)~M~79 was observed on 2010 June 8  as
part of S.T.H.'s fill-in program.

\begin{deluxetable}{ccccccccc}
\tabletypesize{\scriptsize}
\tablewidth{0 pt}
\tablecaption{Swift/UVOT Observations of Target Clusters\label{t:swiftphot}}
\tablehead{
\colhead{Filter} &
\multicolumn{2}{c}{M 67} &
\multicolumn{2}{c}{NGC~188} &
\multicolumn{2}{c}{NGC~2539} &
\multicolumn{2}{c}{M 79} \\
\colhead{} &
\colhead{Exp Time} &
\colhead{Limit} &
\colhead{Exp Time} &
\colhead{Limit} &
\colhead{Exp Time} &
\colhead{Limit} &
\colhead{Exp Time} &
\colhead{Limit)}\\
\colhead{} &
\colhead{(ks)} &
\colhead{(ABmag)} &
\colhead{(ks)} &
\colhead{(ABmag)} &
\colhead{(ks)} &
\colhead{(ABmag)} &
\colhead{(ks)} &
\colhead{(ABmag)}}
\startdata
\hline
$v$    & 0.16 & 18.8 & 2.59 & 20.7 & \nodata & \nodata & \nodata & \nodata \\
$b$    & 0.16 & 19.5 & 2.72 & 21.2 & \nodata & \nodata & \nodata & \nodata \\
$u$    & 0.16 & 19.4 & 2.18 & 22.5 & \nodata & \nodata & \nodata & \nodata \\
$uvw1$ & 2.43 & 21.0 & 3.35 & 22.9 & 1.64    &	20.8   & 2.02    & 22.7\\
$uvm2$ & 2.78 & 21.3 & 6.16 & 23.5 & 1.83    &	21.0   & \nodata & \nodata \\
$uvw2$ & 3.04 & 21.5 & 6.68 & 23.8 & 1.83    &	21.1   & 2.27    & 23.1\\
\hline
\enddata
\end{deluxetable}

\subsection{M~67}
\label{ss:M67}

The open cluster M~67 has been observed extensively from the ground; it is often
used as a calibration standard because it is nearby ($d \simeq 0.8$ kpc), is of
near-solar metallicity, and is relatively old (4~Gyr: VandenBerg \& Stetson
2004; Bellini et al.\ 2010).  However, the only previous published study of M~67
blueward of the atmospheric cutoff at 3000 \AA\ to our knowledge is that of
Landsman et al.\ (1998, hereafter L98), which used data from UIT\null.\footnote{M67
was imaged by the GALEX mission but the complete analysis has not been published
(Landsman et al.\ 2005).} L98
identified twenty UV-bright stars, comprising eleven blue stragglers, seven hot
WDs, one composite yellow-giant\slash WD binary, and one non-member.  However,
the UIT data were not deep enough to delineate the primary sequences (main
sequence, subgiants, giants).

\begin{figure}[ht]
\begin{center}
\includegraphics[scale=.7,angle=270]{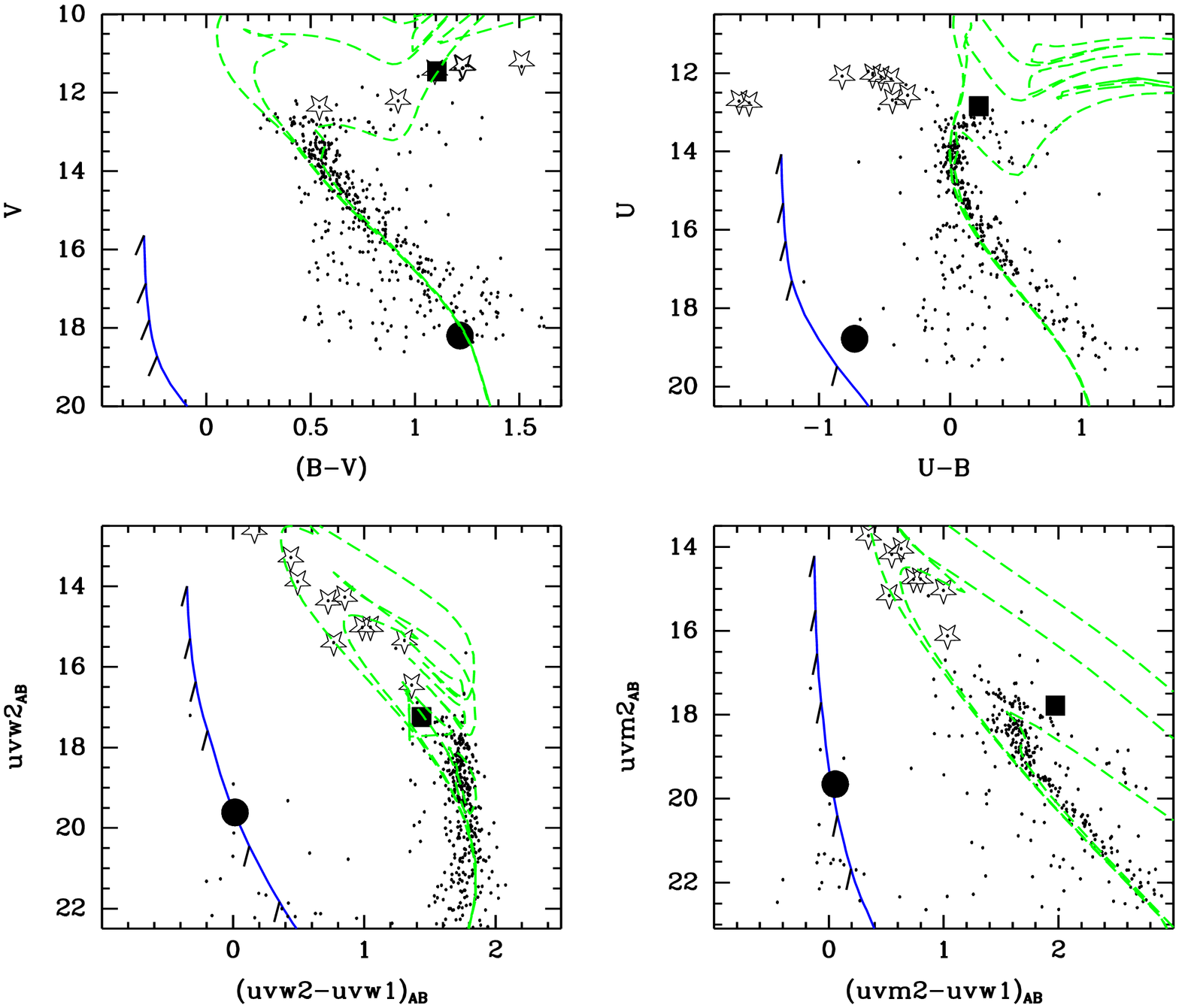} 
\caption{{\it Swift}/UVOT color-magnitude diagrams of M~67. {\it Upper panels:}
optical photometry transformed to Johnson {\it UBV}\null. {\it Bottom panels:}
NUV photometry on AB magnitude system. Dashed lines are isochrones from the
Padova group with $E(B-V) = 0.059$, $(m-M)_0=9.79$, and ages of 500 Myr, 1 Gyr,
and 4 Gyr.  For optical isochrones, we used a metallicity of $\rm[Fe/H]=-0.1$,
but for NUV isochrones we used $\rm[Fe/H]=0$ (see text).  The solid blue line is a
$0.5\,M_{\sun}$ white dwarf cooling curve (Bergeron et al.\ 2011), with tick
marks at temperatures of 100000, 80000, 60000, 40000, 20000, and 10000~K. Open
star symbols are blue stragglers identified in L98; the filled circle is a
potential white dwarf identified in L98; and the filled square is a yellow
binary.\label{f:M67cmd}}
\end{center}
\end{figure}

Figure \ref{f:M67cmd} shows the UVOT optical and NUV CMDs of M~67.   We have
removed objects with DAOPHOT sharpness values greater than 0.5 or less than
$-0.2$ in order to exclude blends, galaxies, and bad measures. The top two
panels show the optical photometry, transformed from $b-v$ and $u-b$ space to
Johnson $B-V$ and $U-B$, respectively, using the formulations of Poole et al.\
(2008), to allow better comparison to previous ground-based photometry and to
theoretical models.  Overlaid with dashed lines are isochrones from the Padova
group (Bressan et al.\ 2012), with a reddening of $E(B-V)=0.059$, and ages of
500 Myr, 1~Gyr, and 4~Gyr, the latter being the nominal age of M~67. The NUV
isochrones include the effects of filter red leaks as well as the non-linear
effect of reddening on the NUV filters (see Appendix A). Providing the best fits to
the isochrones in both the optical and NUV required slightly different
metallicities.  For the optical passbands, we found the best fit was produced
with an age of 4 Gyr, a distance modulus of $(m-M)_0=9.79$, and a metallicity of
$\rm[Fe/H]=-0.1$, all consistent with recent literature (Sarajedini et al.\
2009; Twarog et al.\ 2009;  Pancino et al.\ 2010; Friel et al.\ 2010; Jacobson
et al.\ 2011; Reddy et al.\ 2013). However, for the NUV passbands, we found the
isochrones were better fit with an [Fe/H] of 0.0.  The difference is within the
uncertainties of M~67's metallicity, but may suggest that some modification of
the NUV isochrones is necessary, an issue we will explore in greater detail
later in this series of papers.

We have also overlaid in Figure~\ref{f:M67cmd} a cooling curve for a
$0.5\,M_\sun$ WD from the
compilation\footnote{\url{http://www.astro.umontreal.ca/~bergeron\slash
CoolingModels/}} of P.~Bergeron (Holberg \& Bergeron 2006; Kowaski \& Saumon
2006; Tremblay et al.\ 2011; Bergeron et al.\ 2011).  For the optical filters,
we used the Bergeron synthetic color indices.  For the {\it Swift\/} filters, we
made a simple extrapolation using a black body of the appropriate temperature,
normalized to match the $U$ band synthetic photometry. Future efforts will
refine the WD curves using spectral-energy distributions taken from atmospheric
models of the appropriate temperatures and surface gravities.  The ticks along
the cooling curve mark temperatures of 100000, 80000, 60000, 40000, 20000, and
10000~K from the tip to the faint end. We emphasize that this sequence is simply
overlaid for comparison, and was not fit to any sequence.

NUV colors of red giants in the UVOT filters are strongly affected by filter red
leaks. This produces convoluted isochrones for stars above the subgiant
branch.  However, the main-sequence turnoff and subgiant branch are clearly
defined in all passbands, and correspond roughly to the expectations of the
isochrones.  

Of the eleven blue stragglers identified in L98, nine are within the UVOT field,
and all of them were detected.  A tenth, identified in a previous study of M~67
by Milone \& Latham (1994), is also detected. These are shown in Figure
\ref{f:M67cmd} as open star points.  Note that all are saturated, or close to
saturation, in the three optical filters, creating a spurious roughly flat
distribution in the two optical CMDs. However, in the NUV filters, the blue
stragglers clearly lie along the younger isochrones, consistent with
expectations for merged binary stars.  They also cover a broad range in the NUV
color indices (1--1.5 mag), hinting that the UV has far more sensitivity to the
parameters of blue stragglers than the optical, a concept we will explore in
future contributions.

Of the WDs identified in L98 only one---BATC 2776---falls within the
UVOT field.  This star, marked with a solid circle in Figure \ref{f:M67cmd},
is only marginally detected in the optical, but
has a {\it red} $B-V$ color.  The deep $BV$ photometry Yadav et al.\ (2008) also shows
this object to have a red $B-V$ color of 1.22.   This contrasts against the NUV, where
the object is very blue and sits within a string
of faint hot objects blueward of the main sequence that are not detected in the
optical passbands and lie along the sample WD cooling curve.  They
likely represent the hot end (20-40,000 K) of M~67's WD sequence.

It is possible that this white dwarf is in a binary system with a faint red main sequence star.
The main sequence star shows up faintly in the optical but, because it is cool, disappears in the ultraviolet.
By contrast, the hot white dwarf shows up clearly in the NUV but is too small to be detected in the
optical.  If this is the case, the M67 data confirms the ability of UVOT to detect and
characterize white dwarfs
We detect the tip of the white dwarf sequence and the combination
of NUV and optical data has revealed a previously unknown white dwarf binary.

Finally, the yellow giant with a hot binary companion identified by Landsman et
al.\ (1997)---described as a ``red straggler"---is indicated by a solid square.
In all passbands, we also show it as being slightly redder and brighter than the
MSTO and bluer than the RGB, although the optical photometry is likely
compromised by severe coincidence loss. The companion WD star in M~67's yellow
binary is cooler than the companion star we identified in WeBo 1 (16,000 K vs.\
$\sim$100,000 K; Siegel et al.\ 2012) and is probably close to the limit of
UVOT's ability to detect hidden WD companions.  However, it is notably a little
brighter than the RGB in $uvm2$, which would be consistent with a UV excess.

\subsection{NGC 188}
\label{ss:ngc188}

\begin{figure}[h]
\begin{center}
\includegraphics[scale=.7,angle=270]{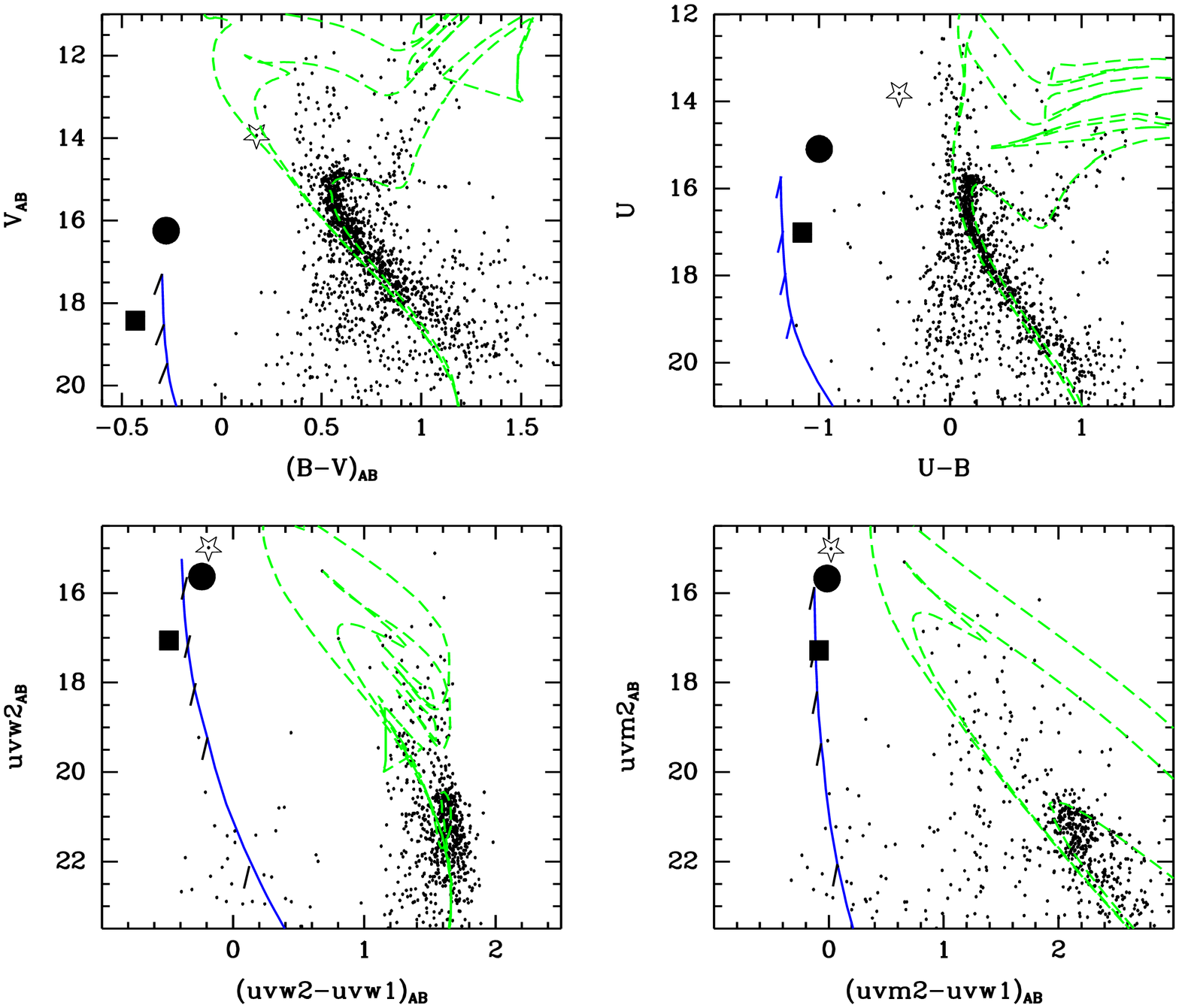}
\caption{NGC~188 color-magnitude diagrams.  Large points are UV-bright stars
identified from UIT data by L98 (see text for details).   Isochrones  from the Padova group are
overlaid with parameters set to [Fe/H]=+0.1, E(B-V) = 0.05, $(m-M)_0$=11.44 and
ages of 500 Myr, 1 Gyr and 5 Gyr.  The solid blue line is a 0.5 $M_{\Sun}$ white
dwarf cooling curve using cooling curves and synthetic photometry from Bergeron
et al.\ (2011). \label{f:ngc188cmd}}
\end{center}
\end{figure}

Figure \ref{f:ngc188cmd} shows the Swift/UVOT CMDs for NGC~188.  We have removed
objects with DAOPHOT sharpness values greater than 0.2 or less than $-0.2$ and
transformed the $b-v$ and $u-b$ photometry to Johnson $B-V$ and $U-B$\null.
Overlaid are isochrones from the Padova group with reddening of $E(B-V)=0.05$, 
ages of 500 Myr, 1 Gyr, and 5 Gyr (the latter being the nominal age of
NGC~188), a metallicity of [Fe/H]=+0.1 and a distance
of $(m-M)$=11.44.  These values are consistent with the literature (Sarajedini et al.\ 1999;
Andreuzzi et al.\ 2002; VandenBerg \& Stetson
2004; Fornal et al.\ 2007; Friel et al.\ 2010; Jacobson et al.\ 2011).
Unlike M~67, we find that a single metallicity of $\rm[Fe/H]=+0.1$ fits
both the optical and NUV sequences. We have also overlaid a cooling curve for a
$0.5\,M_{\sun}$ WD using the synthetic colors of Bergeron et al.\ (2011). The
tick marks along the cooling curve mark temperatures of 100000, 80000, 60000,
40000, 20000, and 10000 K from the tip to the faint end.

L98 identified 7 UV-bright stars in NGC~188 from UIT images and confirmed that
two or three were likely members. Of these, four are too bright for UVOT to
measure and are bright foreground stars (and all but one of these are outside
the field). The three that remain in Figure~\ref{f:ngc188cmd}, marked as large
points, are:

(1) Sandage's (1962) calibration star II-91, for which Dinescu et al.\ (1996)
give a high probability of membership based on its proper motion.  Its spectrum,
according to Green et al.\ (1997), is that of sdB star, and L98 estimate a
temperature of 30,000~K\null.  Our UV photometry (large circle) confirms that this star is UV
bright and is indeed a hot subdwarf.
(2)~Dinescu et al.'s star D-702 is also identified in our sample (starred point) and is also
found to be UV-bright.  L98 note that this star is optically moderately blue
($B-V=0.26$), which we confirm.  We also find that the star is extended in the
images, indicating a possible blend or binary.  The star's photometry bears a
striking resemblance to our recent study of WeBo~1 (Siegel et al.\ 2012) which
we characterized as a red giant with a hot pre-WD companion. 
(3)~UIT-1 shows up faintly in the optical but brightly in the three UV
filters (solid square).  Its extremely blue optical and UV colors are similar to that of the
hot WDs described in Siegel et al.\ (2010).  Its location on the
CMD is near the theoretical WD sequence.

Figure \ref{f:ngc188cmd} shows several other UV-bright stars in NGC~188 that lie
along the younger isochrones.  None have cross-identifications in the Dinescu et
al.\ (1996) catalog, which is not surprising as they lie toward the outskirts of
the cluster.  However, their proximity to the isochrones indicates that they are
likely blue stragglers.  We also find several stars that are optically faint but
detected in the UV and are along the theoretical WD sequences.  These represent
faint WDs or hot subdwarfs.  Further investigation of all of these
stars---especially spectroscopic or astrometric tests of their membership---is
needed to confirm their nature.

\subsection{NGC~2539}
\label{N2539}

\begin{figure}[h]
\begin{center}
\includegraphics[scale=.7,angle=270]{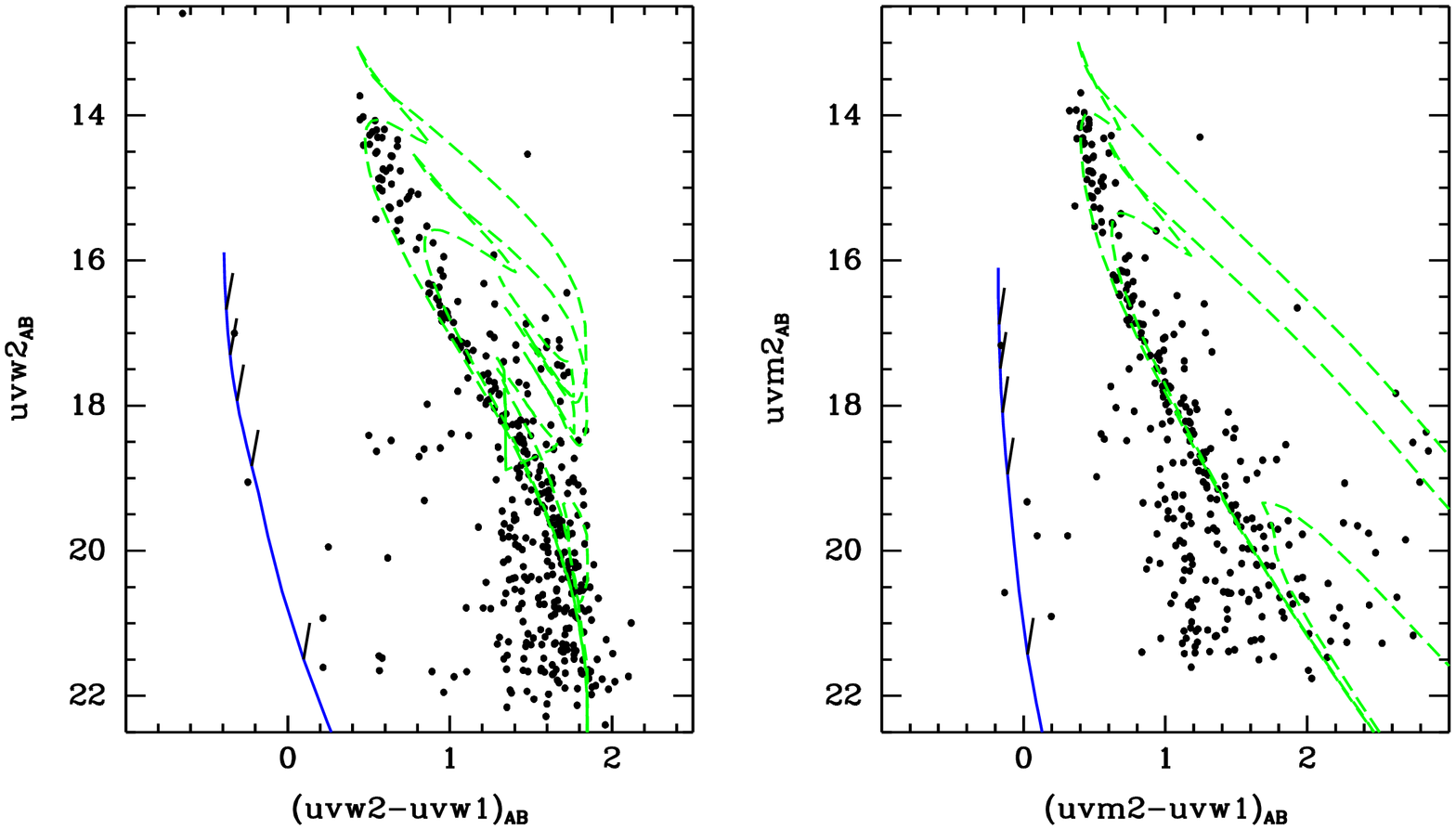}
\end{center}
\caption{NGC~2539 color-magnitude diagrams.  Isochrones  from the Padova group
are overlaid with parameters set to $\rm[Fe/H]=+0.0$, $E(B-V) = 0.05$,
$(m-M)_0=10.65$, and ages of 630 Myr, 1 Gyr and 5 Gyr. The solid blue line is a
$0.5\, M_{\Sun}$ white dwarf cooling curve using cooling curves and synthetic
photometry from Bergeron et al.\ (2011).\label{f:ngc2539cmd}}
\end{figure}

To test the utility of the isochrones for younger populations, we present
photometry of the open cluster NGC~2539. NGC~2539 is of about Hyades age, is
nearby (630 Myr, 1200 pc; Lapasset et al.\ 2000), and of near solar metallicity
(Santos et al.\ 2009; Reddy et al. 2013).  We have overlaid the appropriate
isochrones in Figure \ref{f:ngc2539cmd}, along with a cooling curve for a
$0.7\,M_{\sun}$ WD using the synthetic colors of Bergeron et al.\ (2011). The
tick marks along the cooling curve mark temperatures of 100000, 80000, 60000,
40000, 20000, and 10000 K from the tip to the faint end.

The isochrones reproduce the main sequence very well in both the optical and
NUV, only showing some deviation near the convective hook at the tip of the main
sequence.  Even the evolved sequences in $uvw1-uvw2$ space overlap the Padova
isochrones.  As with M~67 and NGC~188, we show a sequence of objects blueward of
the main sequence that could correspond to extremely hot WDs.  We do not detect
any blue stragglers in NGC~2539, likely due to its young age and the loss of
brighter stars to coincidence saturation.

\subsection{M 79}
\label{M79}

\begin{center}
\begin{figure}[h]
\includegraphics[scale=1]{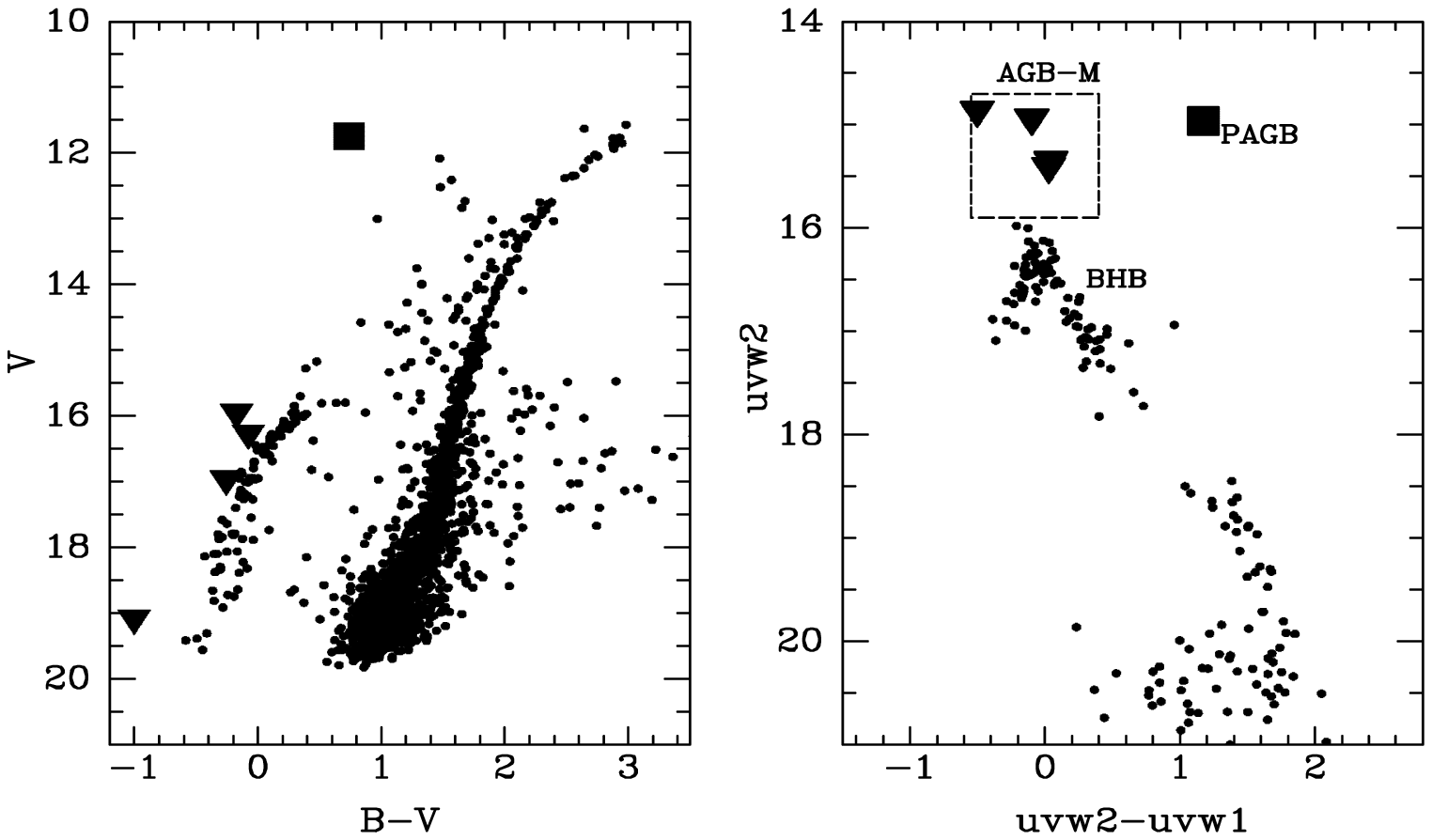}
\caption{M~79 optical and NUV color-magnitude diagrams.  On the left panel, the
PAGB stars is identified with a square while AGB-M stars, as identified from the
UVOT data,  are shown as triangles.  On the right panel, the UV-bright stellar
populations labeled.  M~79's dominant RGB and MS sequences are too faint in the
UV to be detected by Swift/UVOT while the BHB, EHB, and AGB-M stand
out.\label{f:M79cmd}}
\end{figure}
\end{center}

As another test of UVOT's ability to discern hot UV-bright stars, we examine the
old stellar population in a Galactic globular cluster. This population may
contain analogs of the evolved stars that produce the UV upturn in early-type
galaxies.  Figure \ref{f:M79cmd} shows ground-based optical and UVOT NUV CMDs
for the globular cluster M~79. The optical photometry is from frames taken with
the CTIO 0.9-m telescope, obtained by H.E.B. with the Tek3 CCD chip in 1997
November. For the ground-based data, we have restricted our analysis to stars
with DAOPHOT sharp values between $-0.2$ and 0.2.  For the UVOT data, we
retained stars with sharp values between $-0.2$ and 0.5 and lying within $3'$ of
the cluster center.

The optical CMD shows prominent subgiant, RGB, BHB, and AGB sequences.  A hint
of an EHB sequence is seen.  The filled square depicts a known PAGB star ({\c
S}ahin  \& Lambert 2009, initially discovered by Siegel \& Bond (in preparation)
based on ground-based $uBVI$ photometry (see Siegel \& Bond 2005). However, the
NUV CMD is strikingly different from the optical.  The RGB and AGB have
completely vanished, and the BHB, the AGB-M, and the PAGB star are much
more prominent. The triangles in the left panel identify the AGB-M
stars as identified {\it from the UVOT data}.  These were not obviously AGB-M
stars in the optical data (and, in fact, some were too faint or in too heavily
crowded regions to meet our SHARP selection criterion).  This once again
demonstrates the utility of NUV/FUV data in identifying rare evolutionary
sequences.

The point at which the blue end of BHB turns down toward fainter magnitudes in
the NUV data is likely the location of the EHB\null.  For EHB stars, our filters
lose some sensitivity to bolometric correction and being on the Rayleigh-Jeans
tail of the spectral-energy distribution.  We do not detect any clear signature
of blue hook stars in this cluster, but M~79 is not known to  have blue hook
stars and they may be too faint in {\it Swift}'s NUV filters for clear detection
in any case.  We will explore this issue in greater detail with the full data
set, including UVOT photometry of clusters with known blue hook stars.

\section{Conclusions}
\label{s:conc}

Our examination of four nearby star clusters has demonstrated that UVOT 
produces excellent wide-field photometry of stellar populations.  For M~67,
NGC~188, NGC~2539, and M~79, we easily identify unusual UV-bright stars: young
main-sequence stars, blue stragglers, white dwarfs, blue horizontal branch
stars, extreme horizontal branch stars, AGB manqu\'e stars, PAGB stars, and the
hot components of red-giant star\slash white-dwarf binaries and red main-sequence
star\slash white-dwarf binaries.

The main sequences of the young open clusters are consistent with the
expectation of theoretical isochrones, as are the sequences of blue stragglers. 
The young stellar populations show a strong convective hook at the tip of the
main sequence.  UV-bright stars identified in previous surveys---mainly blue
stragglers and white dwarfs---are confirmed by the UVOT data.  While the main
sequence of M~79's old population is too faint in the NUV to be detected, we
clearly show a BHB sequence consistent with previous investigations.  We also
clearly identify the known PAGB star, which stands well away from the primary
sequences.  This indicates that the UVOT will be an excellent tool with which to
find and characterize these extremely rare stars.

Future contributions in this series of papers will examine the complete set of
clusters and nearby galaxies that have been observed by UVOT, in particular the
efficacy of UVOT in separating out composite stellar populations of varying
age and metallicity.  We will also
explore the properties of composite stellar populations in the galaxies of the
Local Group and connect these studies to our ongoing work on unresolved stellar
populations in extragalactic objects. The results with the four test clusters in
the present work, however, confirm the efficacy of our photometric methods and
the ability of UVOT to detect and characterize the UV-bright components of
simple stellar populations.

\acknowledgements

The authors acknowledge sponsorship at PSU by NASA contract NAS5-00136.  This
research was also supported by the NASA ADAP through grants NNX13AI39G and
NNX12AE28G. The authors thank the anonymous referee for useful comments.

\appendix
\section{Correcting Isochrones for Reddening in the UVOT System}

In most photometric analyses, reddening corrections are applied to isochrones
with simple offsets in color-magnitude space scaled to the known reddening
coefficient as tabulated in works such as Schlegel et al.\ (1998).  However,
this simple method may be inappropriate for UVOT data. As demonstrated in Pei
(1992), the UV extinction curve is steep with, for the Milky Way at least, a
large bump at 2175 \AA\null. Additionally, the UVOT filters suffer from some
degree of red sensitivity (Breeveld et al.\ 2011) which complicates the effect
of dust.  This was demonstrated in the context of supernovae by Brown et al.\
(2010b, Section 2.4) and in the context of hot stars by Siegel et al.\ (2012).
Brown's analysis of supernovae showed non-linearity in the UV extinction as
scaled by infrared/optical extinction.  Siegel's analysis of the binary system
WeBo 1 found that for moderately extincted red stars, the effective wavelengths
of the $uvw1$ and $uvw2$ filters, in particular, trended into the visible part
of the spectrum.  This makes a simple linear extinction coefficient problematic
at best.

\begin{center}
\begin{figure}[h]
\includegraphics[scale=1]{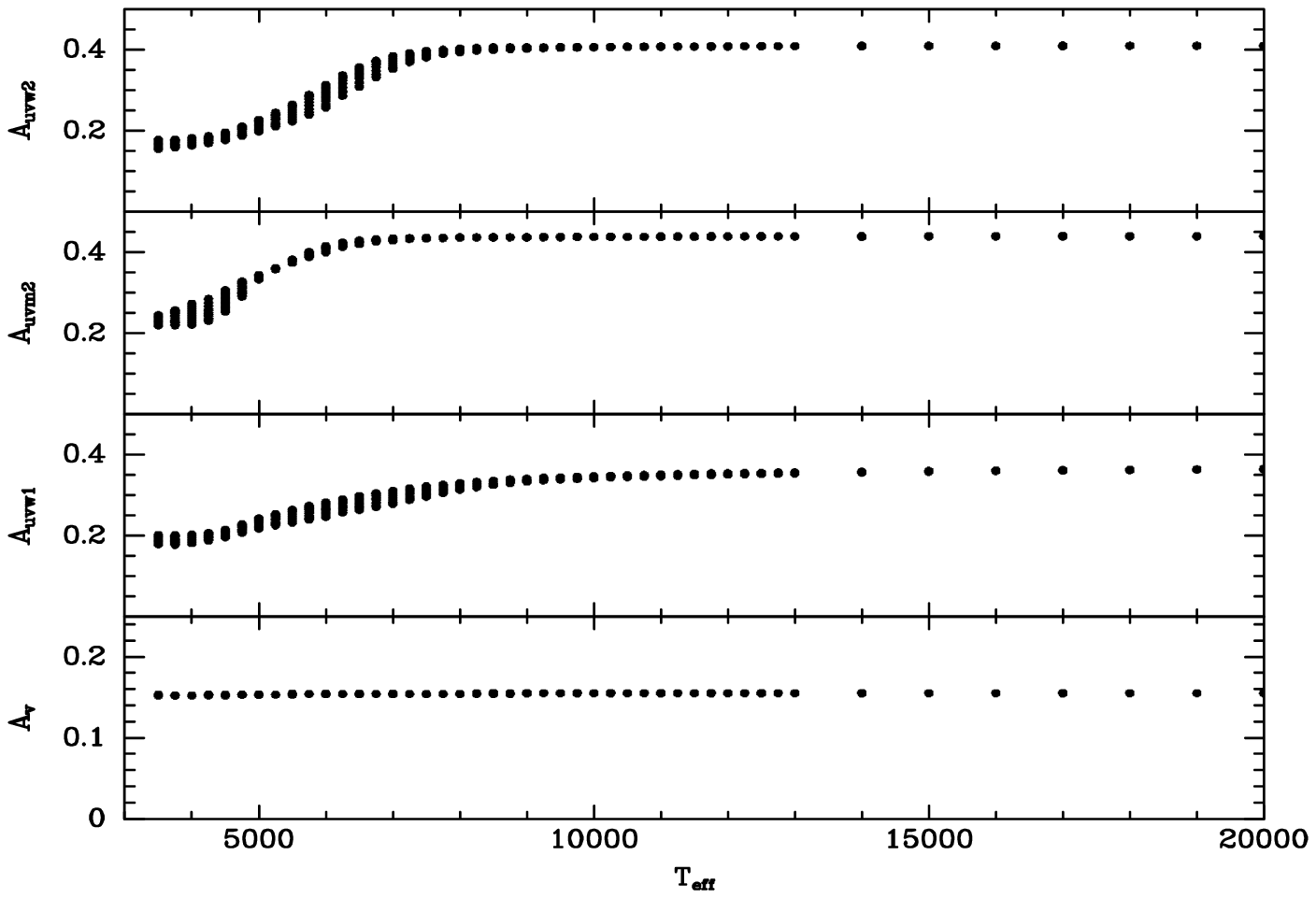}
\caption{The non-linear effect of interstellar reddening in the UVOT filters.  The y-axis shows the magnitude differences between synthetic
photometry generated from the Kurucz models and those generated after the Pei (1992) extinction has been applied with a foreground
reddening of $E(B-V) = 0.05$.  The x-axis shows effective temperature.  Note that cool late type-stars show extinction coefficient
similar to optical filters while early-type stars show much stronger extinction.\label{f:redstar}}
\end{figure}
\end{center}

To test the effects of reddening on the isochrones, we ran all of the solar
metallicity models of Castelli \& Kurucz (2003) through our synthetic photometry
program to generate unreddened magnitudes.  We then reddened the spectra using
the Pei (1992) dust models and re-ran the synthetic photometry code on these
reddened models.  The comparison is shown in Figure \ref{f:redstar} for a
reddening of $E(B-V) = 0.05$.

As can be seen, there is a significant non-linearity in the reddening
coefficients of the UVOT filters.  For cool late-type stars, the UVOT filters
have reddening coefficients similar to optical filters (due to a combination of
red leak and low intrinsic UV flux).   Up to approximately 8,000 K, the
reddening coefficient slowly changes from approximately 3-4 to approximately
6-8, depending on the filter. By comparison, the reddening coefficient in the
$v$ passband is flat with effective temperature. We also find, like Brown et
al.\ (2010), that $A_{\lambda}$ in the UV passbands does not scale linearly with
E(B-V) but declines slightly with increasing reddening.  For example, $A_{uvw2}$
for hot stars declines from 8.17 at E(B-V)=0.1 to 7.80 at E(B-V)=1.0.

We have tabulated the maximum and minimum extinction coefficients for the Swift
UVOT filters---at both low and moderate reddening---in Table
\ref{t:swiftext}.  However, we emphasize that these are the results of
simulation.  Our ongoing programs---especially those that study open clusters
in higher-reddening environments---can provide an empirical constraint on these
values.


\begin{deluxetable}{ccccc}
\tabletypesize{\scriptsize}
\tablewidth{0 pt}
\tablecaption{Extinction Coefficients of Swift UVOT Filters\label{t:swiftext}}
\tablehead{
\colhead{Filter} &
\multicolumn{2}{c}{E(B-V)=0.10} &
\multicolumn{2}{c}{E(B-V)=1.00}\\
\colhead{} &
\colhead{Cool Stars} &
\colhead{Hot Stars} &
\colhead{Cool Stars} &
\colhead{Hot Stars}}
\startdata
\hline
$v$    & 3.04 & 3.10 & 3.03 & 3.09 \\
$b$    & 3.75 & 3.99 & 3.74 & 3.96 \\
$u$    & 4.49 & 4.96 & 4.39 & 4.92 \\
$uvw1$ & 3.54 & 7.40 & 3.38 & 6.76 \\
$uvm2$ & 4.37 & 8.79 & 4.26 & 8.36 \\
$uvw2$ & 3.09 & 8.17 & 2.86 & 7.80\\
\hline
\enddata
\end{deluxetable}

We also note that the use of a single extinction law may itself be problematic. 
It has long been known that the UV extinction law is different in the Large and
Small Magellanic Clouds as contrasted against the Milky Way.  But recent studies
by Ma{\'{\i}}z Apell{\'a}niz \& Rubio (2012), Peek \& Schiminovich (2013), Dong et
al.\ (2014) and Hoversten et al.\ (in prep) have indicated that the UV
extinction law may vary within the same galaxy. There may be significant
variation in the value of $R_V$ and significant variation in the strength of the
2175 \AA\ bump.

Given the multiple dependencies for the UV extinction coefficients---on $A_V$,
on $T_{\rm eff}$ and on the law used---we have ``simplified" things by eschewing
the use of a simple analytical correction.  Instead, the isochrones used
throughout this paper and in future contributions are drawn from a lookup table
which takes the Padova isochrones, finds the nearest match in the Kurucz models
in metallicity-temperature-gravity space, applies the Pei (1992) extinction law
and calculates the reddening offset.  This results in subtle but
not-insignificant changes to the isochrone shape for late-type stars and at high
extinction values.  However, this should have little effect on the hot stars our
survey is focused on as they are sufficiently UV-bright that a near-linear
reddening correction is appropriate.  Further improvements to this program will
account for variations in the reddening curve slope and the strength of the 2175
\AA\ bump as well as the non-linear scaling of the UV reddening with total
line-of-sight extinction.

\clearpage

\bibliographystyle{apj}

\end{document}